\documentclass[12pt]{article}
\usepackage{graphicx}
\usepackage{float}
\begin{document}
\title{Analysis of Lower Hybrid Drift Waves in Kappa Distributions over Solar Atmosphere }
	\author{Antony Soosaleon \& Blession Jose  \\
\small School of Pure \& Applies Physics  \\
\small Mahatma Gandhi University,  
\small Kottayam, Kerala, INDIA -686560\\
\small antonysoosaleon@yahoo.com}
	\date{\today}

\maketitle

\begin{abstract}
Kappa distributions and with loss cone features have been frequently observed with flares emissions with the signatures of Lower hybrid waves. We have analysed the plasma with Kappa distributions and with loss cone features for the drift wave instabilities in perpendicular propagation for Large flare and Normal flare  and Coronal condition . While analysing the growth/damping rate, we understand that the growth of propagation of EM waves increases with kappa distribution index for all the three cases. In comparing the propagation large flare shows lesser growth in compared with the normal and the coronal plasmas. When added the loss cone features to Kappa distributions, we find that the damping of EM wave propagation takes place. The damping rate EM waves is increases with $\frac{T_{\bot}}{T_{\parallel}}$ and loss cone index l, in all the three cases but damping is very high for large flare and then normal in comparision with coronal condition. This shows that the lower hybrid damping may be the source of coronal heating. 
\end{abstract}

\section{Introduction}
Drift instabilities are sometimes also called universal instabilities, a term indicating that drift instabilities are the most general linear instabilities which appear in almost everyplace and at all occasions. The reason is that plasmas are always inhomogeneous at least on the microscopic scales. One may therefore be sure that drift instabilities will be met under all realistic conditions. One of the most important instabilities is the lower- hybrid drift instability. The reason for its importance is that it excites waves near the lower-hybrid frequency which is a natural resonance. Hence, the instability can reach large growth rates. The energy needed to excite the instability is taken from the diamagnetic drift of the plasma in a density gradient \cite{ref1}. This is similar to the modified two stream instability insofar that the diamagnetic drift gives rise to a transverse current in the plasma which acts in a way corresponding to the current drift velocity of the modified two stream instability. In general, the lower-hybrid drift instability is an electromagnetic instability causing whistler waves near the resonance cone to grow \cite{ref1}. 

\subsection{The Kappa Distribution}
Nonthermal particle distributions are ubiquitous at high altitudes in the solar wind and many space plasmas, their presence being widely confirmed by spacecraft measurements \cite{ref4,ref5,ref6,ref7,ref8}. Such deviations from the Maxwellian distributions are also expected to exist in any low-density plasma in the Universe, where binary collisions of charges are sufficiently rare. The suprathermal populations are well described by the so-called Kappa ($\kappa$) or generalized Lorentzian velocity distributions functions (VDFs), as shown for the first time by  Vasyliunas (1968) \cite{ref9}. Such distributions have high energy tails deviated from a Maxwellian and decreasing as a power law in particle speed. Considering the suprathermal particles has important consequences for space plasmas. For instance, an isotropic Kappa distribution (instead of a Maxwellian) in a planetary or stellar exosphere leads to a number density n(r) decreasing as a power law (instead of exponentially) with the radial distance r and a temperature T increasing with the radial distance (instead of being constant).

Scudder (1992a,b) \cite{ref10,ref11} was pioneer to show the consequences of a postulated non thermal distribution in stellar atmospheres and especially the effect of the velocity filtration: the ratio of suprathermal particles over thermal ones increases as a function of altitude in an attraction field. The anti correlation between the temperature and the density of the plasma leads to this natural explanation of velocity filtration for the heating of the corona, without depositing wave or magnetic field energy. Scudder (Scudder 1992b) \cite{ref11} also determined the value of the kappa parameter for different groups of stars. Scudder (1994) \cite{ref12} showed that the excess of Doppler line widths can also be a consequence of non thermal distributions of absorbers and emitters. The excess brightness of the hotter lines can satisfactorily be accounted for by a two-Maxwellian electron distribution function (Ralchenko et al. 2007) \cite{ref13} and should be also by a Kappa. The Kappa distribution is also consistent with mean electron spectra producing hard X-ray emission in some coronal sources (Kasparova and Karlicky 2009) \cite{ref14}. Thus the stability analysis using kappa distribution is of particular interest for solar plasmas. We use a kappa distribution of the form given in Pierrard and Lazar, 2010 \cite{ref15}
\begin{figure}
	\centering
	\includegraphics[width=4.5in]{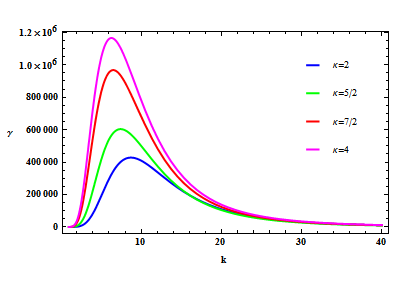}
	\caption{Figure shows plot of growth rate ($\gamma $) versus wave vector (k) for various values of $\,\kappa\,$ index , for  large flares. The growth rate electromagnetic wave in Kappa distribution with loss cone feature shows significant propagation, also the propagation rate increases with kappa index: $\,v_{de}=10^2\,$cm/s $\,T_i=10^7$\,K, $\,T_{\perp}=1.2\times10^7\,$K, $\,T_{\parallel}=10^7\,$K, $\,T_m=10^7\,$K, $\,T_t=2\times10^7\,$K, $\,n_i=n_e=10^{11}\,$cm$^{-3}\,$, B=500 Gauss, $\,\omega=10^8\,$rad/s, $\,v_{de}=~10^{-6}\,$cm/s. \cite{ref2,ref3}}
\end{figure}
\begin{figure}
\centering
\includegraphics[width=4.5in]{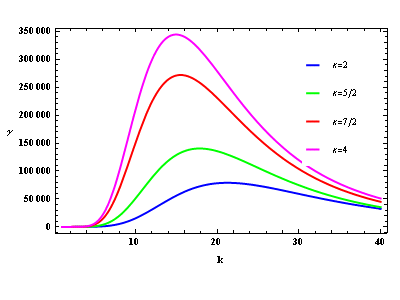}
\caption{Figure shows plot of growth rate ($\gamma $) versus wave vector (k) for various values of $\,\kappa\,$ index , for  normal flares. The growth rate electromagnetic wave shows significant propagation, also the propagation rate increases with kappa index. The growth rate has a wide range of frequencies and  higher in comparision with larger flare and has wide,: $\,v_{de}=10^2\,$cm/s $\,T_i=10^6\,$K, $\,T_{\perp}=1.2\times10^6\,$K, $\,T_{\parallel}=10^6\,$K, $\,T_m=10^6\,$K, $\,T_t=2\times10^6\,$K, $\,n_i=n_e=10^{10}\,$cm$^{-3}\,$, B=300 Gauss, $\,\omega=10^8\,$rad/s, $\,v_{de}=~10^{-6}\,$cm/s.\cite{ref2,ref3}}
\end{figure}\begin{figure}
\centering
\includegraphics[width=4.5in]{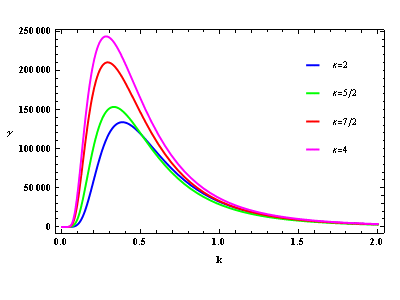}
\caption{Figure shows plot of growth rate ($\gamma $) versus wave vector (k) for various values of $\,\kappa\,$ index , for  coronal condition. The growth rate is higher in comparision with larger but less than normal flare. Corona: $\,v_{de}=10^2\,$cm/s $\,\qquad T_i=10^6\,$K, $\,T_{\perp}=1.2\times10^6\,$K, $\,T_{\parallel}=10^6\,$K, $\,T_m=10^6\,$K, $\,T_t=2\times10^6\,$K, $\,n_i=n_e=10^8\,$cm$^{-3}\,$, B=10 Gauss, $\,\omega=10^6\,$rad/s, $\,v_{de}=~10^{-3}\,$cm/s. \cite{ref2,ref3}.}
\end{figure}
\subsection{Theory}
Out of the large number of waves undergoing quasilinear relaxation we pick out here one particularly interesting electrostatic mode in a magnetized plasma, the lower-hybrid drift mode. Since quasilinear theory describes the time evolution of the equilibrium distribution function, it is necessary to retain the dependence on the distribution function in the expression for the growth rate of the lower-hybrid instability,without assuming it to be a Maxwellian. For the sake of simplicity, we restrict ourselves to purely perpendicular propagation. Then the dispersion relation can be written as \cite{ref1} 
\begin{equation}
D(k_{\perp},\omega)=1+ \chi_e(k_{\perp},\omega)+\chi_i(k_{\perp},\omega)
\label{Eq.8.1}
\end{equation}
where $\,\chi_e(k_{\perp},\omega)=\frac{\omega_{pe}^{2}}{\omega_{ge}^{2}}(1-\frac{1}{k_{\perp}L_R}\frac{\omega_{ge}}{\omega-k_{\perp}v_{de}})\,$ and $\,\chi_i(k_{\perp},\omega)=\frac{\omega_{pe}^{2}}{k_{\perp}}\int \frac{dv_{\perp}}{\omega-k_{\perp}v_{de}}\frac{\partial{f_{0i}(v_{\perp},t)}}{\partial{v_{\perp}}}\,$ where $\,\omega_{pe}\,$ and $\,\omega_{ge}\,$ are the plasma and gyro frequency respectively.
The growth rate in the low drift velocity regime $\,(v_{de} < v_{thi})\,$, is,

\begin{equation}
\gamma=-\frac{\pi v_{thi}^{2}k_{\perp}^{2} v_{de}}{2 |k_{perp}| (1+\frac{k_{\perp}^{2}}{k_{max}^{2}})}\frac{\partial{f_{0i}}}{\partial{v_{\perp}}}\|_{v_{\perp}\rightarrow \omega/ k_{\perp}}
\label{Eq.8.2}
\end{equation}
Here $\,v_{thi}=\sqrt{\frac{2 k_B T_i}{m_i}}\,$ is the thermal velocity of  protons  and $\,v_{de}\,$ is the drift velocity of electrons.
It is assumed that $\,v_{de}=-v_{di}\,$ and $\,k_{max}\,$ defined as,

\begin{equation}
k_{max}^{2} \lambda_{Di}^{2}=2/(1+\omega_{pe}^{2}/\omega_{ge}^{2})
\label{Eq.8.3}
\end{equation}
\begin{figure}
	\centering
\includegraphics[width=4.5in]{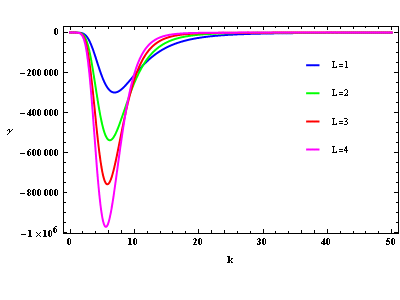}
	\caption{The figure shows the damping rate of Electromagnetic waves for various values of L  for large flares. We find the damping increases with the loss cone index, this signifies the loss of particles increases the damping rate, which may be the source of heating mechanism. Same datas as for figure 1}
\end{figure}
\begin{figure}
	\centering
\includegraphics[width=4.5in]{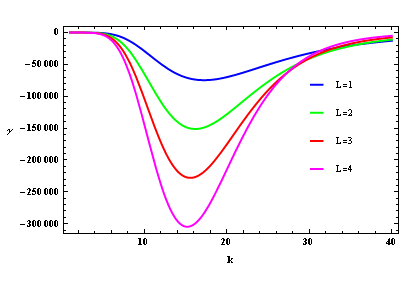}
	\caption{The damping rate of EM wave with wave vector graph at various values of L  for  normal flares . The damping rate is lesser than the large flare but wider range of frqquency is observed,  Same datas as for figure 2}
\end{figure}
\begin{figure}
	\centering
\includegraphics[width=4.5in]{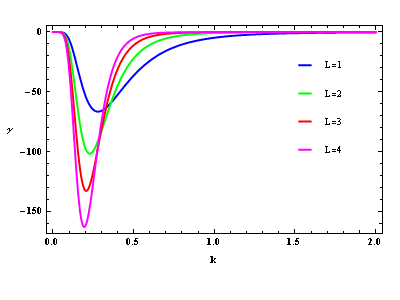}
	\caption{The damping rate for coronal plasma for at various values of L  for  corona. The damping is much less compared with both the flares plot.Same datas as for figure 3}
\end{figure}

where $\,\omega_{pe}\,$ and $\,\omega_{ge}\,$ are the plasma and gyro frequencies respectively.

\begin{equation}
f=\frac{n}{2\pi\,(\kappa \,w_{\kappa i}^{2})^{3/2}} \frac{\Gamma(\kappa+1)}{\Gamma(\kappa-1/2)\,\Gamma(3/2)}\left(1+\frac{v_{\parallel}^{2}+v_{\perp}^{2}}{\kappa w_{\kappa i}^{2}}\right)^{-(\kappa+1)}
\label{Eq.8.8}
\end{equation}

where $\,w_{\kappa i}=\sqrt{(2\,\kappa-3)\,k_B T_i/\kappa \,m_i}\,$ is the thermal velocity.

Substituting this in the general equation for the growth rate Eq.(2) we get the growth rate of lower hybrid drift wave corresponding to the kappa distribution as,

\begin{equation}
\gamma_{\kappa}=\frac{-\pi\,v_{thi}^{2}k_{\perp}v_{de}}{\left(1+\frac{k_{\perp}^{2}}{k_{max}^{2}}\right)^2} \frac{n_i\,\Gamma(\kappa+1)(-1-\kappa)}{\Gamma(\kappa-1/2)\Gamma(3/2)\,\kappa\,w_{\kappa i}^{2}}\frac{\omega}{k}\left[1+\frac{2}{\kappa} \frac{\omega^2}{k^2 w_{\kappa i}^{2}}\right]^{-2-\kappa}
\label{Eq.8.9}
\end{equation}

The instabilities are studied for large flares, medium flares and corona. Drift velocities $\,v_{de}=10^{-14}\,$cm/s, $10^{-13}\,$ cm/s and $10^{-10}\,$ cm/s taken for large flares, medium strength flares and corona respectively. .

\subsection{Kappa Distribution with a Loss Cone Feature}

A distribution function of this feature is described in Summers and Throne (1995) \cite{ref16} and Singhal and Tripathi, 2006 \cite{ref17}.
It has the following form
\begin{equation}
\begin{array}{r}
f=\frac{\Gamma(\kappa +l+1)}{\pi^{3/2}\theta_{\perp i}^{2}\,\theta_{\parallel i}\, \kappa^{l+3/2}\Gamma (l+1)\Gamma (\kappa - 1/2)}(\frac{v_{\perp}}{\theta_{\perp i}})^{2l} \times \\ \\ \left[ 1+\frac{v_{\parallel}^{2}}{\kappa \theta_{\parallel i}^{2}}+\frac{v_{\perp}^{2}}{\kappa \theta_{\perp i}^{2}} \right]^{-(\kappa +l+1)}
\end{array}
\label{Eq.8.10}
\end{equation}

where, $\,\theta_{\perp i}=(\frac{2\kappa-3}{\kappa})^{1/2}(l+1)^{-1/2}(\frac{T_{\perp i}}{mi})^{1/2}\,$ and $\,\theta_{\parallel i}=(\frac{2\kappa-3}{\kappa})^{1/2}(\frac{T_{\parallel i}}{m_i})^{1/2}\,$.
\section{Results and Discussions}

Substituting Eq. 6 in the Eq. 2 , the general expression for the growth rate and assuming $\,|v_{\parallel}\ = |v_{\perp}| =\omega/k\,$ we get the growth rate as

\begin{equation}
\begin{array}{l}
\gamma = -\frac{\pi}{2} \frac{v_{thi}^{2}k_{\perp}^{2} v_{de}\,\Gamma(\kappa +l+1)}{2|k_{\perp}| \,(1+\frac{k_{\perp}^{2}}{k_{max}^{2}})^{2}}   \frac{1}{\pi^{3/2}\theta_{\perp i}^{2}\theta_{\parallel i}\, \kappa^{l+3/2}\Gamma(l+1)\Gamma(\kappa -1/2)} \times \\ \lbrace \frac{2(-1-\kappa -l)(\frac{\omega}{k_{\perp}})(\frac{\omega}{k_{\perp}\,\theta_{\perp i}})^{2l}(1+\frac{\omega^2}{\kappa k_{\perp}^{2}\theta_{\parallel i}^{2}}+\frac{\omega^2}{\kappa k_{\perp}^{2}\theta_{\perp i}^{2}})^{-2-\kappa -l}}{\kappa \theta_{\perp i}^{2}} + \\ \frac{2l(\frac{\omega}{k_{\perp}\theta_{\perp i}})^{-1+2l}(1+\frac{\omega^2}{\kappa k_{\perp}^{2}\theta_{\parallel i}^{2}}+\frac{\omega^2}{\kappa k_{\perp}^{2}\theta_{\perp i}^{2}})^{-1-\kappa-l}}{\theta_{\perp i}} \rbrace
\end{array}
\label{Eq.8.11}
\end{equation}

The growth/damping rates are studied for large and normal flares and corona for drift speeds $\,v_{de}=10^2\,$cm/s have been analysed. We are finding that the propagation of EM waves are very prominant In the thermal relation graphs $\,\kappa\,$ is fixed at a value $5/2$.

\begin{figure}
	\centering
	\includegraphics[width=4.5in]{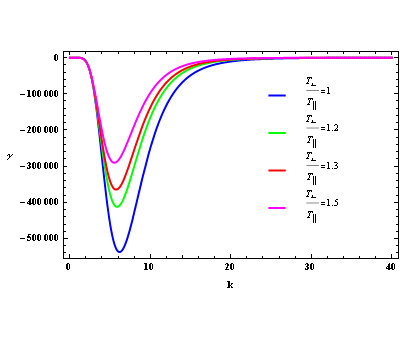}
	\caption{The growth rate wave vector graph at various values of thermal ratios for  large flares, Same datas as for figure 1 }
\end{figure}
\begin{figure}
	\centering
	\includegraphics[width=4.5in]{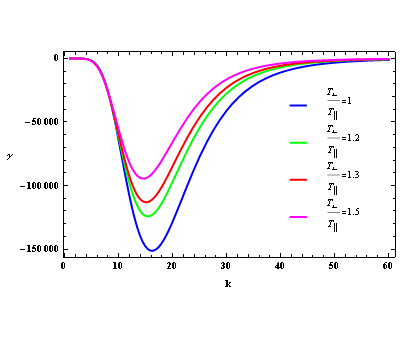}
	\caption{The growth rate wave vector graph at various values of thermal ratios for  normal flares Same datas as for figure 2}
\end{figure}

\begin{figure}
	\centering
		\includegraphics[width=4.5in]{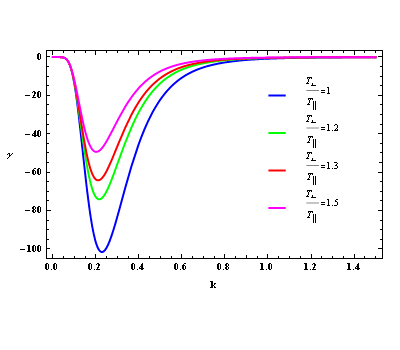}
		\caption{The growth rate wave vector graph at various values of thermal ratios for corona, Same datas as for figure 3}
\end{figure}

\section{Conclusion}

We have analysed the instability of perpendicular propagating lower hybrid drift waves in the solar atmosphere. The growth/ damping in the low drift velocity regime is analysed for  the Kappa distribution with a loss cone feature. We see a common characteristic for all the cases. The growth/ damping rate decreases as the perpendicular to parallel thermal ratio of ions $\,(\frac{T_{\perp}}{T_{\parallel}})\,$ increases. The stability analysis performed for kappa distribution for ambient corona, medium sized flares and large flares. The growth rate increases with increase in $\,\kappa\,$ values and is independent of $\,\frac{T_{\perp}}{T_{\parallel}}\,$. Space plasmas exhibit a distribution described by these features in the waves, and if this were the case, what would be the instability characteristic of the lower hybrid drift waves. The results shows wave damping, the highest damping rate being for large solar flares. The damping increases with increase in the index $L$. The damping rate also increases with increase in $\,\frac{T_{\perp}}{T_{\parallel}}\,$ which may be the source of coronal heating.


\end{document}